# FULL TITLE

# Energy Dissipation Modulates Sensitivity to Corner Advantage in Weighted Dice

Nick Kinney[1,*], Karissa Boyce[1], Hallie Nelson[1], Stephen Loftus[1], Lem Curran[1], Joanna Friesen[1]

[1] Sweet Briar College, 134 Chapel Road, Sweet Briar, VA 24595

* To whom correspondence should be addressed. Tel: 301-338-1181

Email: nkinney@sbc.edu

# Abstract

Weighted dice have pedagogical value and a range of broader implications. The physics of such dice provides an entry point for understanding state probabilities, center-of-mass calculations, and inertia tensors. Related considerations arise in the physics of grains, seed dispersal, erosion, and materials science, all of which involve components that can be compared with the stochastic tumbling and settling of dice. Previous empirical results are consistent with a Boltzmann-like distribution. We reproduce this result at scale using 23 classes of corner-loaded dice and more than 500,000 physical dice rolls collected with an automated rolling and imaging system. Briefly, landing probability increases exponentially with corner advantage (lower center-of-mass). We test the additional hypothesis that landing statistics depend on the combined effects of corner advantage, inertia tensor, and energy dissipation. Corner advantage defines the direction of the gravitational preference, but the rate of energy dissipation determines how strongly that preference is expressed in the final landing distribution. Simulations suggest that the inertia tensor and its orientation affect roll statistics; indeed, we find a small but statistically detectable departure from uniform landing probabilities under our experimental conditions. Our approach has pedagogical value as an entry-level engineering and data science project. The dice are 3D-printed and automatically rolled at scale using affordable parts and open-source tools. Results are compared to rigid-body simulations and a six-state model of energy dissipation.

# Introduction

The statistical behavior of weighted dice is relevant to systems in which objects tumble stochastically and eventually settle into one of several discrete orientations. Examples include the physics of grains, the dispersal of seeds, erosion, and the motion of manufactured components[1–4]. These systems are far more complicated than dice; nevertheless, a weighted cube provides a controlled experimental model in which geometry, mass distribution, gravitational energy, and rotational inertia can be varied independently or in combination. A foundational experimental study of loaded dice was performed by Levin, who examined seven plastic cubes containing asymmetric brass loads[5]. Levin described each resting orientation as an energy state and analyzed the observed frequencies using Boltzmann-like expressions involving gravitational potential energy, activation energies for tumbling, and the number of tumbling channels. Each die was tossed at least 600 times, and the resulting orientation frequencies showed broad agreement with these energy-based models[5].

Center-of-mass position is not the only property changed by an internal load. Altering the distribution of mass also changes the inertia tensor, which determines the resistance of the body to angular acceleration about different axes[6,7]. A die with unequal principal moments of inertia (inertial anisotropy) may tumble differently depending on the magnitude of those moments and their alignment with die-face normal vectors. In particular, an internal loading scheme can produce a preferred principal axis associated with the smallest moment of inertia. If the dynamics of tumbling and impact retain information about this axis, two dice with similar centers of mass but different inertia tensors could exhibit different landing probabilities. Conversely, if energy dissipation and repeated collisions erase this dynamical information, landing statistics may depend primarily on gravitational potential energy.

Here, we reproduce and extend the empirical investigation of loaded-dice statistics using 3D printing, automated rolling, and computer vision. We compare 23 rotation-distinct classes of corner-loaded dice that vary in center-of-mass, total added mass, and inertial anisotropy. An automated apparatus repeatedly rolls multiple dice, photographs each outcome, and identifies the upper face using machine-readable fiducial markers. This system produced more than 500,000 physical rolls, permitting comparisons among dice classes, individual physical replicates, and face-specific physical predictors. Roll dynamics such as face transitions and energy dissipation are studied using high-speed video. Results are compared with rigid-body simulations and a simple 6-state model of energy dissipation.

We test two related hypotheses. First, we ask whether the observed face probabilities follow the gravitational-energy dependence reported in earlier loaded-dice experiments. Second, we ask whether landing probabilities contain an additional contribution from the inertia tensor, its orientation, and rates of energy dissipation between dice classes. These effects are evaluated using physical calculations of the center-of-mass and inertia tensor for each loading geometry. Our results reproduce the previously reported energy-dependent behavior and provide evidence that the inertia tensor has a small but detectable effect on final face probabilities under our experimental conditions. Thus, dice with markedly different internal mass distributions can have similar roll statistics when those distributions produce equivalent gravitational relationships among the faces. On the other hand, the rate of energy dissipation may modify landing bias. Heavier dice may settle more rapidly while exploring a similar number of face transitions compared with lighter dice; consequently, heavier dice are less sensitive to their corner-loading geometry.

Beyond its physical conclusions, the experiment provides a scalable instructional framework. The dice can be manufactured with consumer 3D printers, and the rolling apparatus can be assembled from inexpensive electronic and mechanical components. Image collection and analysis can be performed using open-source software. The project therefore integrates mechanics, probability, statistical modeling, computer vision, automation, and reproducible data analysis in a system whose physical behavior is directly observable.

# Results

## Automatic, reproducible, and scalable collection of independent dice rolls

The goal of this study is to understand how center-of-mass, inertia tensor, and energy dissipation affect the roll statistics of weighted dice. Previous studies suggest that center-of-mass effects dominate the face roll statistics of weighted dice but do not take into account the additional factors considered here. In particular, the principal moment of inertia is smallest about the minimum-principal-moment axis (Fig. 1A); therefore, we hypothesized that the relative ease of angular acceleration about this axis might influence landing statistics. In addition, energy dissipation may modify center-of-mass effects. We designed 23 classes of weighted dice to study the combination of these effects. Each class carries a unique set of corner loadings with $2^8$ combinations reduced to 23 by rotational symmetry (Fig. 1A). Dice classes were 3D printed, and the large number of collected rolls was well suited to analysis (Fig. 1A, inset).

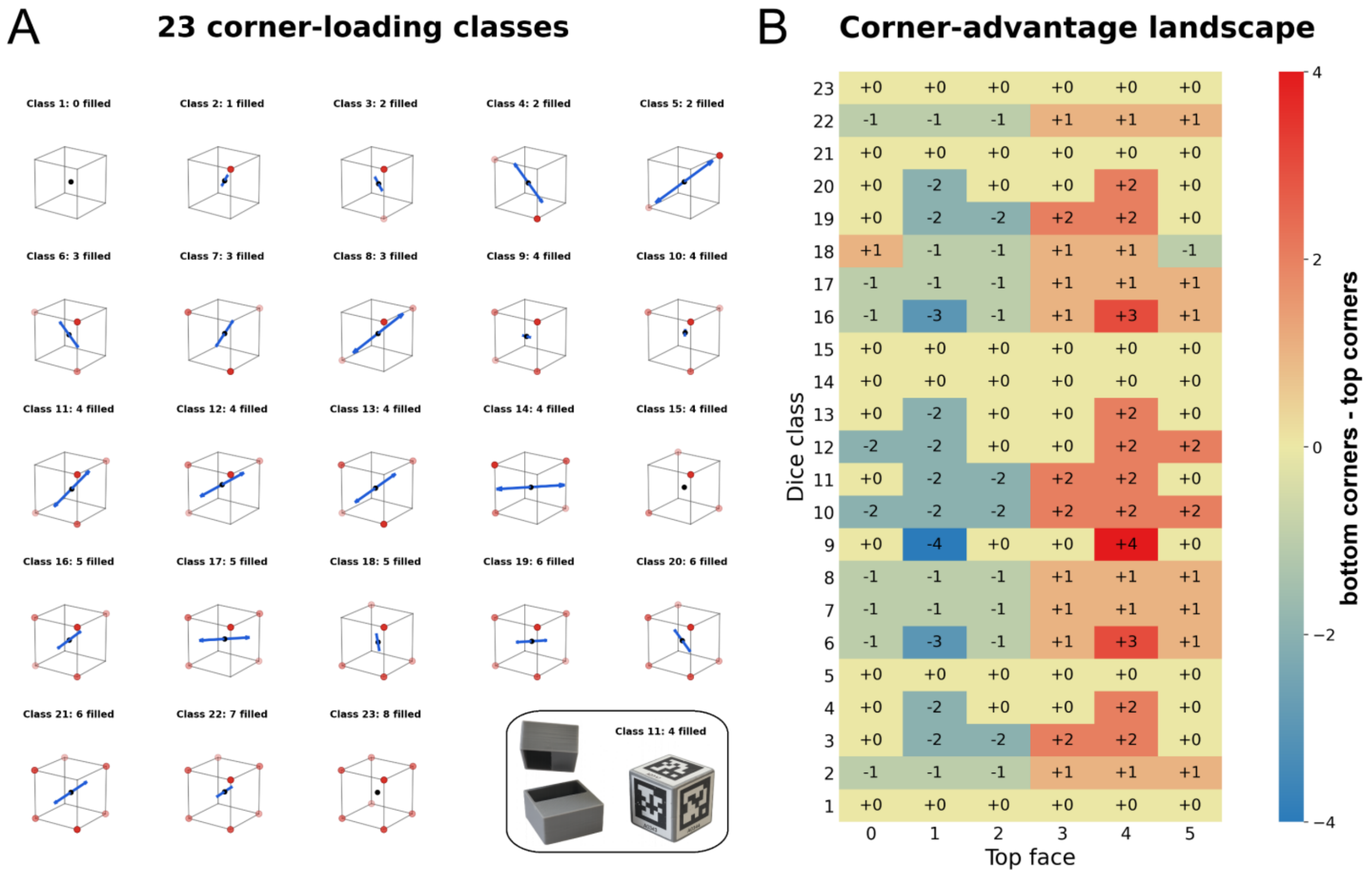

**Figure 1. Structural diversity of the 23 corner-loading classes. (A)** Geometric representations of the 23 distinct corner-loading configurations. Red markers indicate filled corners, the black point marks the center-of-mass, and the blue line indicates the minimum principal axis of inertia. **(B)** Corner-advantage landscape for each die class and possible top face. Values give the number of filled corners below the die minus the number above. Positive values indicate configurations with more filled corners below the die, negative values indicate more filled corners above it. Inset ArUco code reprinted from [12] under a CC BY license, with permission from Rafael Salinas, original copyright 2014.

Dice rolls were collected at scale using a seesaw attached to a servo motor synchronized to a computer vision system (see methods for details). Briefly, the seesaw consists of three parts: a randomizer, a ramp, and a landing zone. Control dice (Fig. 1A, classes 1 and 23) were used to ensure roll independence. This approach enabled us to collect more than 500,000 rolls in a two-week period, improving on previous studies by several orders of magnitude[5]. The system also has pedagogical value because it combines 3D printing, computer vision, automation, and data analysis in a system whose physical behavior is directly observable.

## Internal mass loading weakens the relationship between corner advantage and landing probability

Previous experiments have shown that the outcome frequencies of physically asymmetric dice can be represented by Boltzmann-like functions of gravitational energy[5]. Levin used such a model to describe seven cubes loaded with asymmetrically positioned brass inserts, while later work applied a related “synthetic temperature” formalism to more than 50,000 rolls of dimensionally asymmetric dice[8]. We therefore asked whether a comparable relationship would emerge for dice whose external geometry remained constant but whose internal mass distributions varied systematically. For each possible upper face *f* of die class *c*, it is useful to consider the corner advantage:

$$\Delta_{cf} = N_{bottom,cf} - N_{top,cf}$$

Here, $N_{bottom,cf}$ and $N_{top,cf}$ are the numbers of filled corners adjacent to the lower (resting) face and upper (visible) face, respectively. Positive values correspond to orientations in which more of the added mass lies below the geometric center of the die (Fig. 1B). Across 515,754 rolls, landing probability increased approximately exponentially with corner advantage (Fig. 2A). We represented this relationship using the normalized Boltzmann-like model

$$P(f|c) = e^{\beta_n \Delta_{cf}} \Big/ \sum_{j=1}^{6} e^{\beta_n \Delta_{cj}}$$

where $\beta_n$ describes the sensitivity of landing probability to corner advantage among dice with *n* filled corners. The fitted value of $\beta_n$ decreased monotonically as total corner filling increased, from approximately 0.300 for dice with two filled corners to 0.163 for dice with six filled corners. Thus, increasing internal mass loading did not simply magnify the effect of corner placement. Instead, the landing distribution became progressively less sensitive to each additional unit of bottom-versus-top mass asymmetry. Across the five fitted values, an exact one-sided Spearman rank test showed a significant negative monotonic association between total corner filling and $\beta_n$ ($\rho=-1.00$, $p=0.0083$).

Corner advantage alone did not fully determine the probability of a face. The size and structure of the dataset allowed us to resolve smaller effects that would likely be obscured in a more limited experiment. In particular, faces with the same corner advantage did not always land with the same probability. For faces with a corner advantage of zero, landing probability declined as the corner advantages of the two neighboring axes increased (Fig. 2B). Thus, the probability of a face was determined not only by its own energetic advantage, but also by the surrounding energy landscape. This result follows naturally from a Boltzmann-like description of the landing probabilities. Let the resting face have corner advantage A (the visible face will have corner advantage −A). The four neighboring faces form two additional opposite pairs with advantages ±B and ±C. The probability of the resting face is

$$e^{\beta A} \Big/ 2[\cosh(\beta A) + \cosh(\beta B) + \cosh(\beta C)]$$

where $\beta$ is the fitted effective energetic sensitivity. When $A$=0, increasing either $B$ or $C$ increases the denominator and therefore decreases the probability of the resting face. The apparent neighbor effect therefore does not necessarily require an additional force or bias; it arises because Boltzmann probabilities are normalized over the complete set of possible landing states rather than being determined by the landing face in isolation. The neighbor-axis landscape also has a mechanical interpretation. Changing the distribution of mass along the neighboring axes changes the rotational resistance associated with tipping away from that face. We quantified this contribution using the mean in-plane moment,

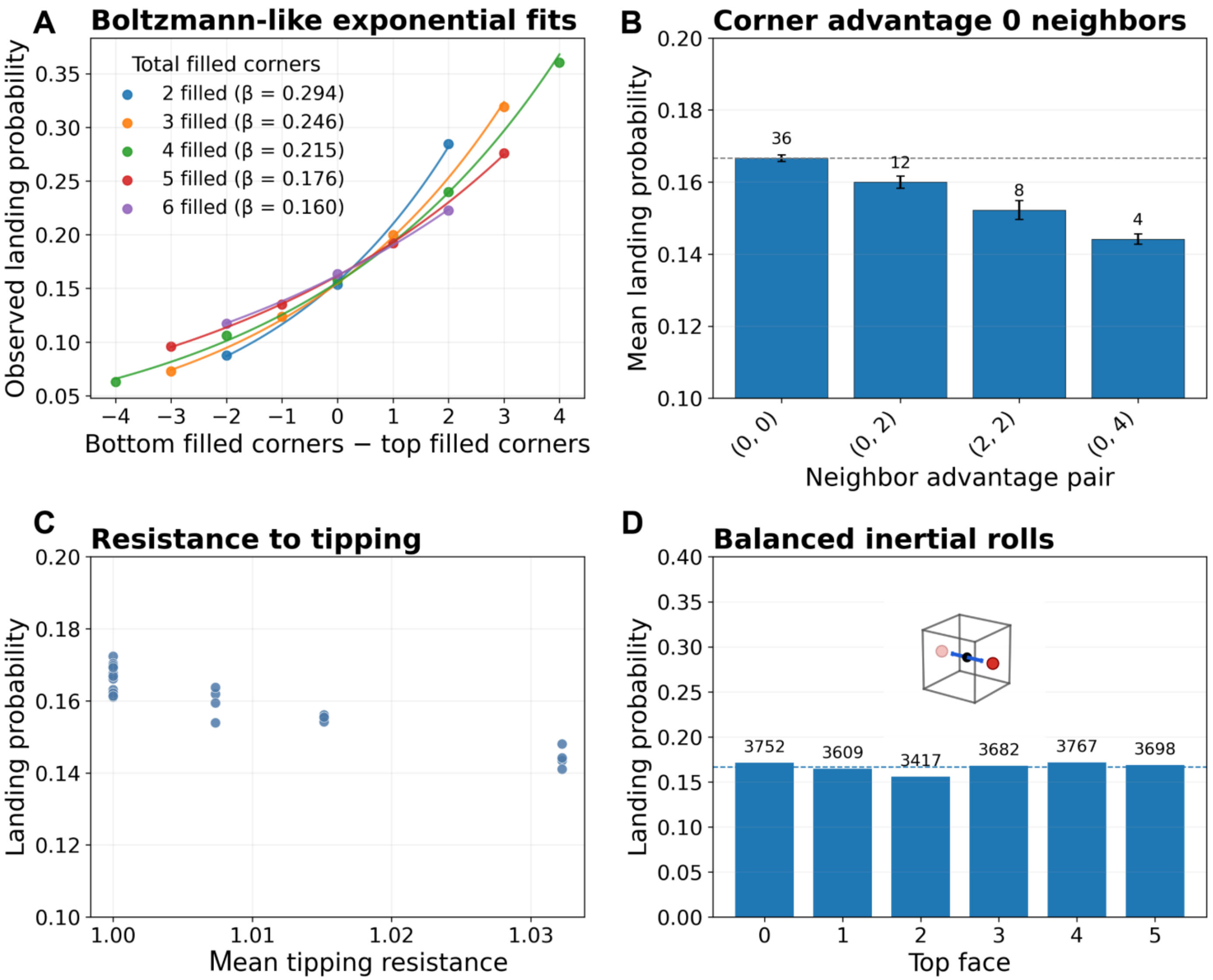

**Figure 2. Gravitational asymmetry and rotational accessibility jointly shape the landing probabilities of internally weighted dice. (A)** Mean observed landing probability as a function of corner advantage for dice containing two through six filled corners. Lines show fitted Boltzmann-like exponential relationships, with fitted sensitivity parameter β. **(B)** Mean landing probability of faces with a corner advantage of zero, grouped by the corner advantages along the two neighboring axes. Numbers above the bars indicate the number of faces represented in each group; the dashed line indicates the fair-die probability of 1/6. **(C)** Observed landing probability as a function of mean tipping resistance for faces of four-corner-loaded dice with a corner advantage of zero. **(D)** Pooled face probabilities from 21,925 rolls of a balanced but anisotropic die. The inset shows the die geometry, face-centered masses, center-of-mass, and minimum principal axis of inertia.

$$I_{tip} = {tr(\boldsymbol{I}) - \boldsymbol{n}^T \boldsymbol{I} \boldsymbol{n}}/_2$$

where **n** is the landed-face normal and **I** is the moment of inertia tensor. Greater calculated resistance to tipping is associated with a lower landing probability (Spearman's ρ = -0.985, p = 0.0056, Fig. 2C). Because both neighbor advantage and tipping resistance arise from the same underlying mass geometry, this relationship should not be interpreted as an independent causal effect. Nevertheless, it motivated a more direct examination of whether the inertia tensor itself contributes detectably to landing statistics. The inertia-tensor representations in Figure 1 show the minimum-principal-moment ($I_{min}$) for each die class as blue arrows indicating the axis of rotation. Their length represents the dimensionless anisotropy

$$A = {I_{max} - I_{min}}/_{[(I_1 + I_2 + I_3)/3]}$$

Thus, a "large" blue axis represents a large separation among the principal moments, not a large absolute value of $I_{min}$.

The original 23 corner-loaded classes did not provide a clean example in which a strongly anisotropic minimum-moment axis was aligned with a face normal. We therefore introduced an additional, 24th class (Fig. 2D, inset) containing equal masses at the centers of two opposite faces. This design kept the center-of-mass fixed while generating a pronounced minimum-principal-moment axis, thereby isolating rotational anisotropy from the gravitational bias produced by corner loading. The pooled face distribution showed a small but statistically detectable departure from the fair-die expectation of 1/6 ($\chi^2$=11.61, p=0.040). The observed pattern was consistent with a slight reduction in landing probability for faces whose normals were aligned with the minimum-moment axis. Under the present rolling conditions, the inertia tensor therefore produced a detectable but small effect compared with the much larger effects of the center-of-mass energy landscape.

## MuJoCo simulations reproduce the empirical bias but do not support a simple rotational-energy partition explanation

To test whether the empirical landing probabilities could be reproduced from rigid-body dynamics alone, we simulated the 23 corner-filled dice classes in MuJoCo[9]. The simulation used the same corner-loading geometry as the physical dice and recorded the stable upward face after each die came to rest. Despite the simplified environment, the simulated rolls reproduced the central empirical pattern observed in the physical data: for dice with 2–6 filled corners, landing probability increased approximately exponentially with corner advantage (Fig. 3A). As in the experimental data, the fitted Boltzmann-like curves became shallower as the number of filled corners increases, suggesting that dice with more filled corners are less sensitive to the same corner advantage. Interestingly, MuJoCo simulations of the inertially anisotropic die (class 24) predicted increased landing probabilities for the two oppositely loaded, center-of-mass-balanced faces (Fig. S2). The discrepancy from experiment is beyond the scope of this study but may reflect idealized contact and friction parameters, the point-mass approximation used to represent the internal weights, or differences between the simulated and physical release and landing conditions.

The simulations also reproduced a more subtle feature of the physical system regarding the effect of neighboring faces (Fig. 2B). When the corner advantage is zero, landing probabilities also depended on the distribution of filled corners along the two neighboring axes (Fig. 3B). Despite deviation from experiment for the (0, 4) neighbor-axis advantage pair, this result indicates that the model is also sensitive to the broader three-dimensional mass distribution of the die. The presence of this effect in simulation supports the use of MuJoCo as a useful mechanistic approximation of the experimental system. Interestingly, the MuJoCo simulations suggest that dice with more filled corners tend to visit more faces before settling (Fig. 3C). In addition, the rotational kinetic-energy fraction differs by corner-fill class during the simulated settling process (Fig. 3D).

The simulation results suggest that heavier, more corner-filled dice may undergo a different relaxation process, visiting more faces before reaching a stable resting state. This additional tumbling could partially dilute the effect of any single face's energetic advantage: a favored orientation may be reached, lost, and revisited multiple times before the die finally settles. In this interpretation, the shallower Boltzmann-like fits for heavier dice reflect a higher effective dynamical “temperature” and greater mixing during settling. The final landing probability still depends on corner advantage, but that bias is filtered through a longer sequence of stochastic face transitions. These simulations therefore play two roles. First, they show that the Boltzmann-like landing curves and neighbor-axis effects can emerge from rigid-body dynamics with realistic mass distributions. Second, they argue against a simple static energy-partition mechanism as the main reason that heavier dice show reduced corner-advantage sensitivity. This motivated analysis of high-speed video to directly count face transitions in real dice as they tumbled to rest.

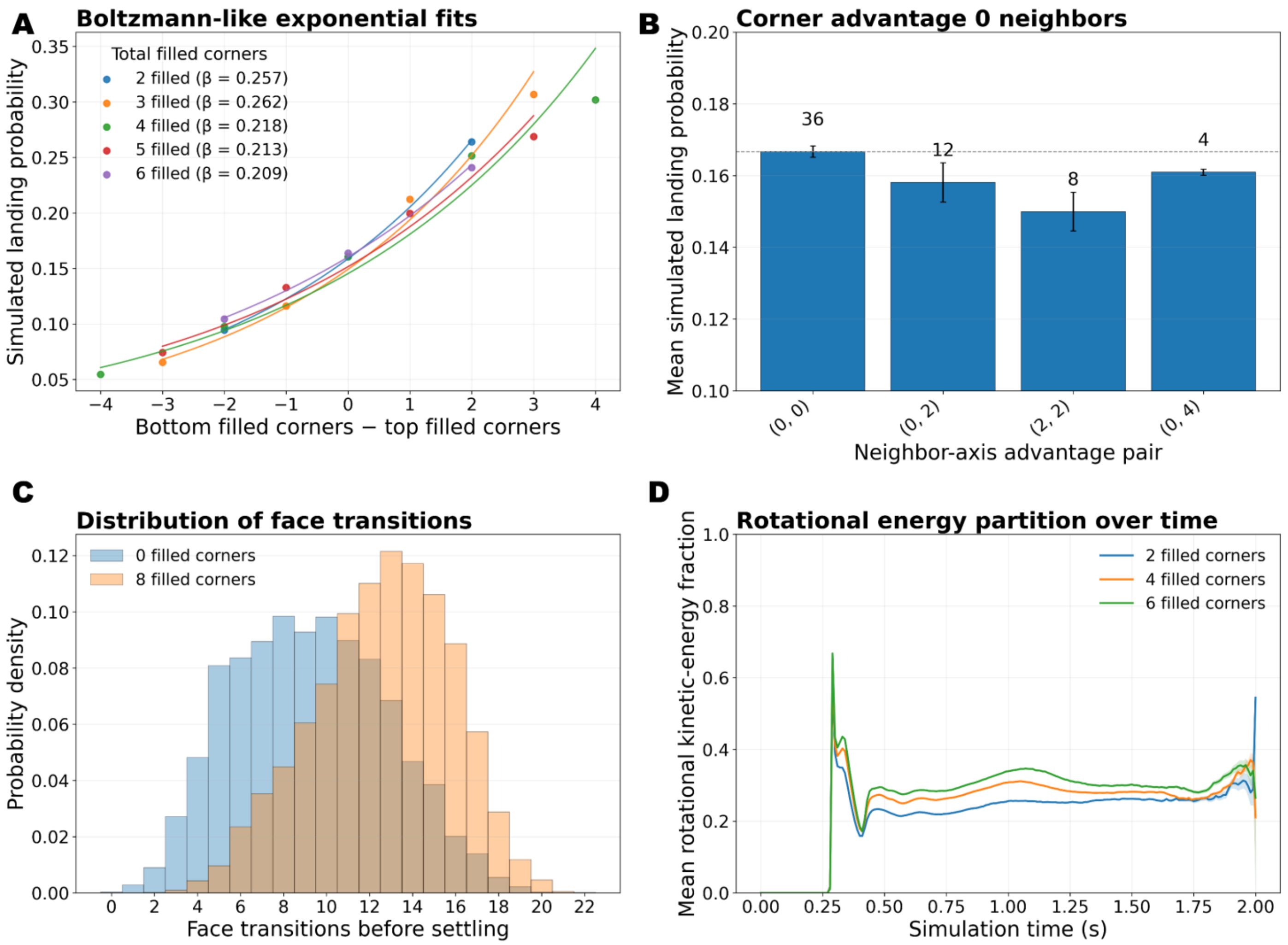

**Figure 3. MuJoCo simulations reproduce Boltzmann-like landing biases but do not support a simple rotational-energy partition mechanism.** Rigid-body simulations were performed in MuJoCo using dice models with the same corner-fill classes as the physical dice. **(A)** Simulated landing probabilities for dice with 2–6 filled corners increase approximately exponentially with corner advantage, reproducing the Boltzmann-like relationship observed experimentally. **(B)** Among simulated faces with zero corner advantage, landing probability varies with the distribution of filled corners along neighboring axes, indicating that the broader three-dimensional mass distribution contributes to landing behavior. The dashed line marks the fair-die expectation of 1/6. **(C)** Simulated dice with more filled corners tend to undergo more face transitions before settling. **(D)** The simulated rotational kinetic-energy fraction over time is similar for different dice classes. Together, these simulations suggest that the reduced corner-advantage sensitivity of heavier dice is not explained by a lower rotational-energy partition alone, but may instead reflect differences in energy dissipation and greater mixing among face states before settling.

## Fully corner-loaded dice settle more rapidly despite similar numbers of face transitions

To test whether the simulated differences in rotational dynamics could be detected experimentally, we recorded high-speed videos of rolls from the two extreme corner-loading configurations: unfilled dice with no added corner masses and fully corner-loaded dice with all eight corners filled. A total of 116 rolls were analyzed for each class. The die was detected in each video frame, the visible top face was classified, and face assignments were used to identify face transitions. Settling was defined as the beginning of the final interval in which the detected die remained stationary.

The total number of detected face transitions was broadly similar between the two loading classes ($p=0.4627$, Fig. 4A). Thus, adding corner mass did not produce a large change in the tally of face transitions. Transition rates also overlapped strongly between the two groups ($p=0.0816$, Fig. 4B). Most rolls from both classes underwent approximately 10–25 transitions per second. The similarity of the transition-rate distributions also indicated that the fully loaded dice did not simply rotate more slowly because of their larger moment of inertia.

A clearer difference was observed in the time required to settle ($p=2.27e-05$, Fig. 4C). The fully corner-loaded dice were concentrated at shorter settling times, with most rolls reaching their final stationary state within approximately 0.5–1.0 s after first detection. In contrast, the unfilled dice more frequently remained active for 1.0–1.5 s and produced a broader right tail of long-duration rolls. The detected centroid path lengths were broadly similar between the two groups (p=0.0121, Fig. 4D).

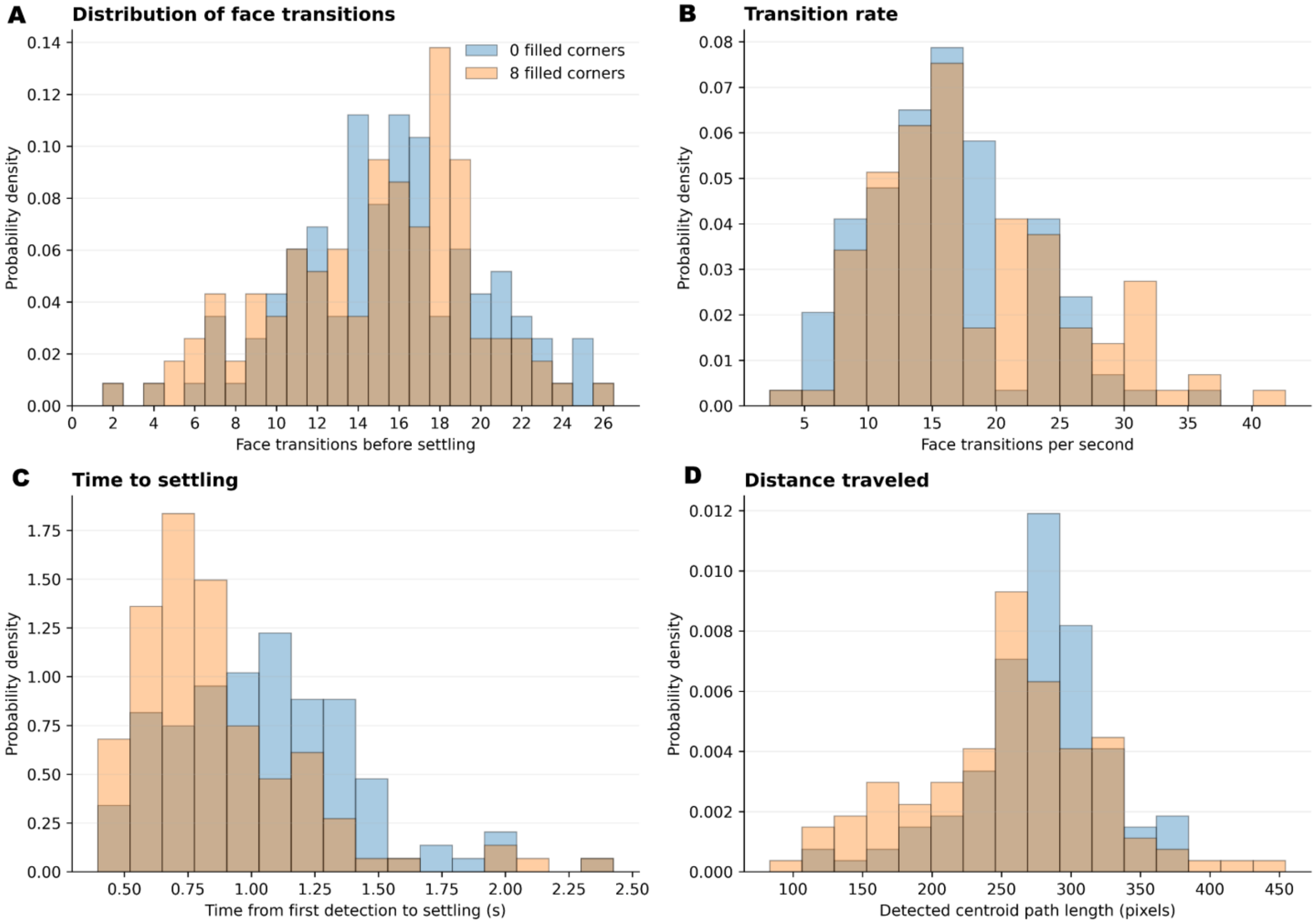

**Figure 4. Experimental rolling dynamics of unfilled and fully corner-loaded dice.** Distributions are shown for 116 rolls of dice with zero filled corners and 116 rolls of dice with all eight corners filled. **(A)** Total number of confirmed face transitions detected in each video. **(B)** Face-transition rate, calculated relative to the measured settling interval. **(C)** Time from the first reliable die detection to the beginning of the final settled interval. **(D)** Total path length of the detected bounding-box centroid before settling. Histograms are normalized to probability density.

Although the experimental dynamics differed from the MuJoCo predictions, the fully corner-loaded dice settled more rapidly than the unfilled dice while undergoing a similar number of detected face transitions. On the other hand, the simulations predicted more transitions for the fully loaded class (Fig. 3C). Together, these observations are consistent with faster loss of usable mechanical energy in the fully loaded dice; however, energy dissipation was not measured directly.

## A minimal six-state model links energy dissipation to sensitivity to corner advantage

To clarify how dissipation rate could affect corner-loading bias, we constructed a minimal six-state model representing the six possible upper faces of a three-corner-loaded die. The states were assigned upper-face filled-corner counts of 3, 2, 2, 1, 1, and 0 (Fig. 5A). Each trajectory began in a uniformly random state. At each step, one adjacent state change (Δs) was proposed while the available mechanical energy of the system (E(t)) decayed exponentially. Downhill or equal-energy transitions were accepted, whereas uphill transitions were accepted only when their energetic cost was less than that of the system. A trajectory ended when the remaining energy fell below the activation energy required for another transition ($E_{hop}$).

Trajectories of the model – simulated using python – show that the distribution of resting states depends strongly on the dissipation rate (Fig. 5B). At low dissipation rates, the system has longer to explore states and is less likely to become trapped in unfavorable resting states. Consequently, the resting state with corner advantage +3 was favored in the final outcome. As the dissipation rate increased, the system was more likely to become trapped in unfavorable resting states below the activation energy. At high dissipation rates the system quickly dropped below the activation energy and kept its original random state. Consequently, all states were equally probable.

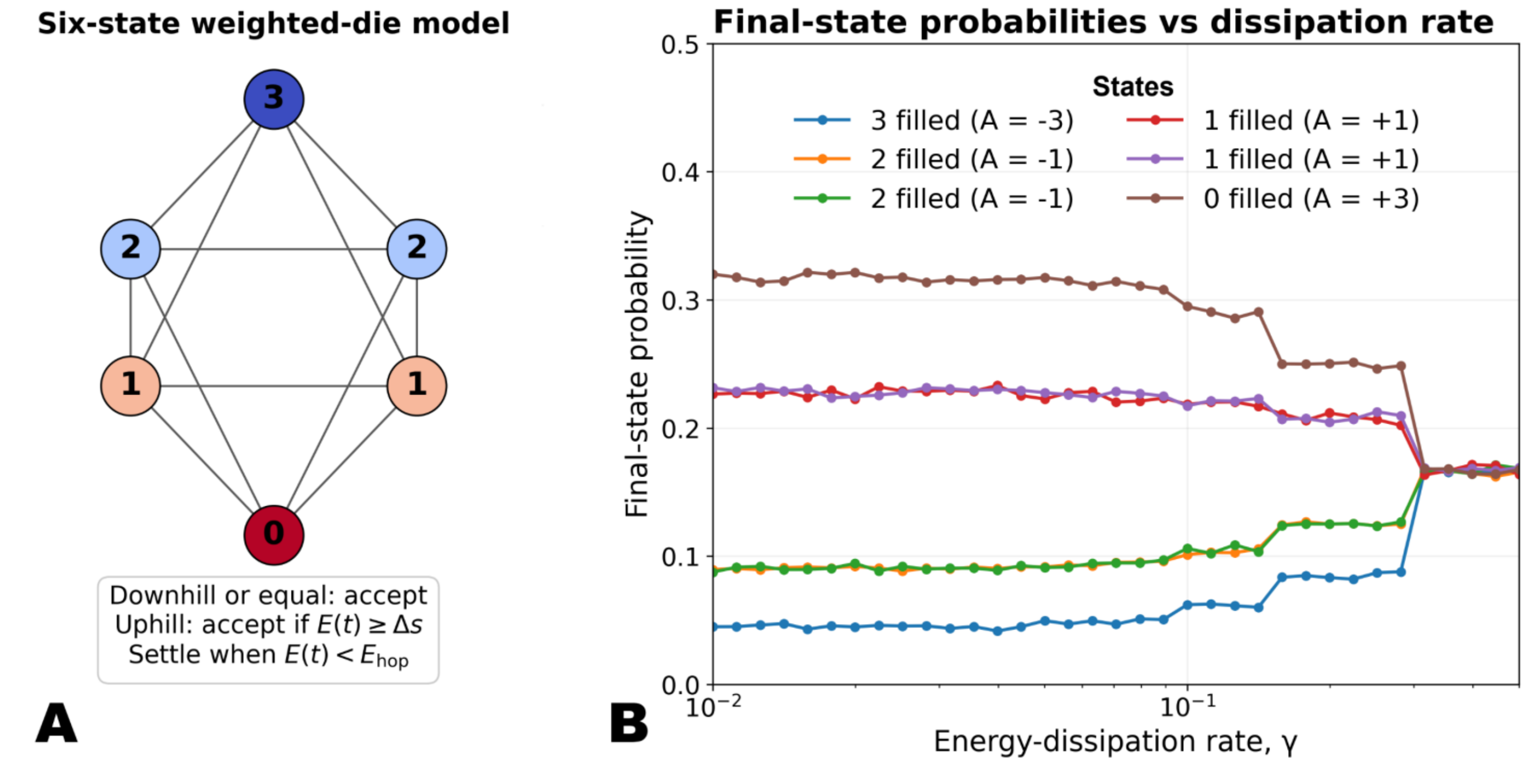

**Figure 5. Energy dissipation controls sensitivity to corner advantage in a six-state weighted-die model. (A)** Network of the six possible resting faces of a three-corner-loaded die. Nodes show the number of filled corners on the upper face, and edges connect adjacent faces. Proposed transitions are accepted according to the difference in corner advantage (Δs), activation energy ($E_{hop}$), and the system's energy (E(t)). **(B)** Final-state probabilities as a function of dissipation rate, γ. Faster dissipation limits exploration of the state network and leaves the final-state distribution closer to the uniform initial distribution.

This conceptual model illustrates the central conclusion of the study: **corner advantage defines the direction of the gravitational preference, but the rate of energy dissipation determines how strongly that preference is expressed in the final landing distribution.** The fitted Boltzmann-like sensitivity parameter observed for physical dice can therefore be interpreted in part as a measure of how rapidly a tumbling system loses the mechanical energy required to escape from favorable orientations. The six-state model is intentionally simplified; it provides a mechanistic link between energy dissipation and the experimentally observed sensitivity to corner loading.

# Discussion

This study shows that the outcomes of internally weighted dice are governed by both the gravitational landscape of the resting states and the dynamics through which the die loses mobility. Across 23 distinct corner-loading classes and more than 500,000 physical rolls, landing probability increased approximately exponentially with corner advantage, confirming that orientations placing more added mass below the geometric center are favored. However, the fitted sensitivity to corner advantage decreased as total corner loading increased. Neighboring-face geometry produced additional variation among faces with the same corner advantage, while the balanced but strongly anisotropic die showed a small but statistically detectable departure from uniformity. Thus, inertial anisotropy can modestly affect landing probabilities under these conditions, although its effect was much weaker than the biases produced by center-of-mass displacement.

Our experimental findings are restricted to a single die geometry, rolling surface, and obstacle layout which necessitate a discussion of their robustness. We consider each main conclusion. First a Boltzmann-like distribution is reproduced at scale. This result was originally demonstrated by Levin who reported tossing die a minimum of 600 times[5]. That’s a very different rolling mechanism demonstrating the same distribution with orders of magnitude fewer rolls. This leads us to believe that a Boltzmann-like distribution may be robust to rolling details. Second, the balanced anisotropic die (class 24) deviates from uniform landing probabilities. Simulations also show deviation but differ in magnitude and direction (Fig. S2). Thus, effects of the inertia tensor may be sensitive to the aforementioned details; however, this does not invalidate the conclusion that those effects are measurable and open to further investigation. Third, rate of energy dissipation determines how strongly corner advantage affects the final landing distribution. This conclusion is supported by the minimal six-state model and to some extent seems trivial. Instantaneous energy dissipation in the six-state model results in a uniform face distribution. One could go further and speculate that landing probabilities depend on the type of dissipation; for example, monotonic dissipation vs exponential dissipation. Exploring these possibilities is beyond the scope of this work.

Previous studies established that loaded dice produce Boltzmann-like outcome distributions[5,8], but the present work extends those observations in several ways. The external geometry was held constant while the internal mass distribution was varied systematically across the complete set of rotation-distinct corner-loading patterns. The scale of the automated dataset also permitted smaller effects—such as the influence of the surrounding six-state landscape—to be resolved. Most importantly, the combination of physical rolls, rigid-body simulations, and high-speed trajectory analysis allowed gravitational preference, rotational geometry, and dissipative settling to be examined separately. The fitted parameter β should therefore not be interpreted solely as a measure of center-of-mass displacement or as an effective temperature. Instead, it is a measure of how strongly an energetic landscape is expressed under a particular rolling and dissipation protocol.

This interpretation may be relevant to other systems in which irregular objects tumble and eventually settle into one of several stable orientations. Grains transported by wind or water, seeds moving across soil, fragments undergoing erosion, and manufactured components moving through sorting or assembly processes all combine deterministic mechanics with highly variable collision histories. In such systems, geometry determines the available stable states, but the observed state distribution may also depend on how rapidly translational and rotational energy are removed. A rapidly dissipating object may become trapped before fully exploring its orientation landscape, whereas a more persistent object may repeatedly leave and revisit favorable states. The dice therefore provide a tractable example of a broader nonequilibrium principle: final-state probabilities reflect both the relative preference of the states and the timescale over which transitions remain dynamically accessible.

The system also has substantial pedagogical value[10]. In fact, previous studies have used weighted dice to introduce concepts such as entropy, chemical energy, and Gibbs free energy in a classroom setting[11]. Our dice are fabricated with consumer 3D printers, rolled automatically using an Arduino-controlled apparatus, and analyzed with affordable cameras and open-source software. The project connects center-of-mass calculations, inertia tensors, probability distributions, computer vision, Monte Carlo modeling, rigid-body simulation, and reproducible data analysis within a physical experiment whose outcomes are immediately understandable. Students can formulate predictions from the loading geometry, collect large datasets, compare experiment with simulation, and evaluate where simplified models succeed or fail. The modular design also permits projects of different complexity, ranging from elementary comparisons of face frequencies to advanced analyses of high-speed trajectories and nonequilibrium state transitions.

Further work should measure energy dissipation more directly and independently of corner loading. Three-dimensional tracking could be used to estimate translational and rotational kinetic-energy decay, while high-speed measurements across intermediate loading classes could test whether settling time changes systematically with the fitted value of $\beta$. Dissipation could also be manipulated without changing mass geometry by varying surface material, friction, restitution, ramp angle, or die compliance. The six-state model could be expanded to include measured initial-state distributions, state-specific transition barriers, finite transition times, and collision-driven energy changes. Such experiments would clarify how much of the observed decline in corner-advantage sensitivity is explained by dissipation itself and how much reflects other correlated consequences of increasing mass. More generally, they would help establish corner-loaded dice as a reproducible benchmark for connecting rigid-body dynamics, dissipative trapping, and probabilistic outcomes.

# Methods

## Automatic roller design

An automated dice roller was used to collect a large number of reproducible, approximately independent rolls (Fig. 6A). The apparatus used a servo motor to drive a seesaw style rolling platform synchronized with a camera that detected ArUco markers assigned to individual die faces[12]. The platform was constructed from a standard 18 × 18 × 16-in. cardboard box with one side removed. The interior surfaces of the remaining panels formed the rolling track, which consisted of three sections: a randomizer, a ramp, and a landing zone. The ramp and landing zone were lined with a continuous piece of felt to provide a conventional dice-rolling surface. The original fold between these sections was fixed at an interior angle of 150°, such that dice entered the landing zone from a 30° incline. The dimensions of the ramp and landing zone allowed most, although not all, dice to come to rest before image acquisition and face detection. All code used for the project is freely available online: nkinney06/corner_loaded_dice.

The randomizer was constructed as a removable modular component from the remaining cardboard panel (Fig. 6A). To minimize the moving mass of the seesaw, lightweight rhombus-shaped weighing boats were fastened to the cardboard surface and supplemented with Styrofoam obstacles. These structures introduced repeated collisions and changes in direction before the dice reached the ramp and landing zone, thereby reducing dependence on their initial positions and orientations. Although the precise degree of randomization produced by individual obstacles was not quantified, the control dice, represented by classes 1 and 23, produced approximately uniform face frequencies and showed no detectable serial dependence between successive rolls. These controls indicate that the apparatus did not itself introduce a strong persistent face bias or roll-to-roll memory.

The measured landing probabilities may nevertheless depend on features of the apparatus, including the ramp angle, rolling-surface material, obstacle geometry, release conditions, and the translational and rotational velocities of the dice as they entered the landing zone. We therefore do not assume that the numerical values of fitted parameters, including the effective inverse-temperature parameter, are universal or independent of the rolling protocol. Instead, the principal result is the reproducible agreement between landing probability and a Boltzmann-like function of the gravitational energy landscape across many distinct mass configurations. Apparatus-specific conditions may alter the effective scale of this relationship by changing the amount of kinetic energy available during settling, but the consistency of its functional form across die classes suggests that the Boltzmann-like description is robust to substantial variation in mass distribution and initial rolling trajectory. This interpretation does not establish equilibrium sampling in a strict thermodynamic sense; rather, it identifies an effective statistical description of the outcomes generated by the present rolling system.

## Design and fabrication of 23 classes of corner loaded dice

The design of the 23 corner classes of weighted dice began by considering all $2^8$ combinations of corner-fill. We then applied the cube's 24 rotational symmetries (6 faces * 4 orientations each) to the 256 patterns, yielding the 23 rotation distinct classes. Dice were 3D-printed in two halves (Fig. 6B) using standard PLA filament before gluing the two halves together and applying ArUco stickers to each of the six faces. An STL file suitable for 3D printing is available as supplementary material: four copies are needed for a complete set of dice.

Each die class was fabricated in triplicate not only for reproducibility but also to control for potential asymmetry introduced when the two halves of a die are glued together (Fig. 6C). Our results do not indicate that the glue plane affects roll statistics of any of the 23 classes of dice.

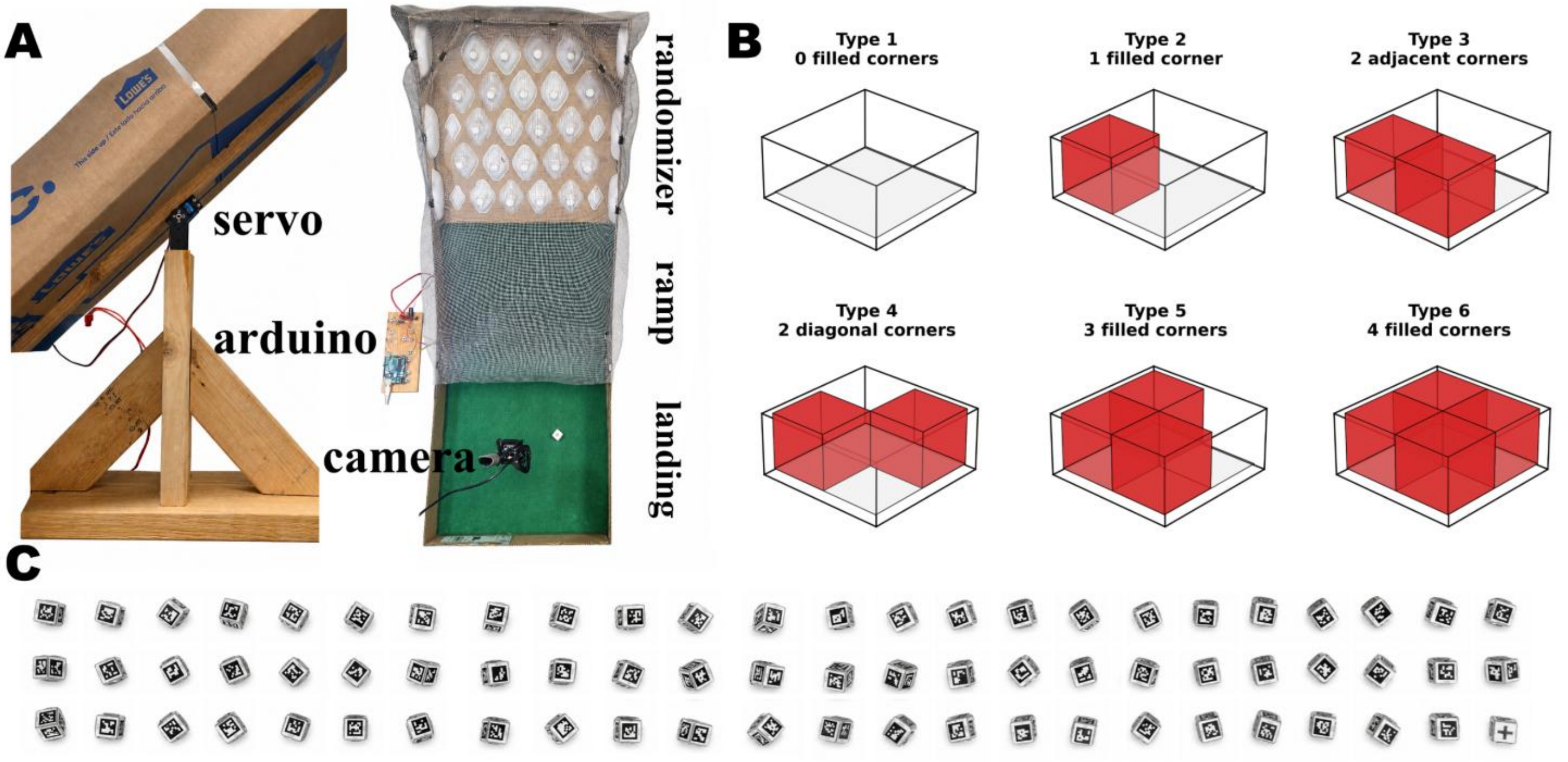

**Figure 6 – Automatic roller design and construction of 23 classes of corner loaded dice. (A)** An automatic dice roller uses an Arduino to drive a servo motor that tilts a seesaw-style rolling platform synchronized to a camera. The camera detects ArUco markers assigned to individual die faces. **(B)** Dice are 3D printed in two halves before gluing and applying ArUco stickers to each face. Each die class was rolled in triplicate using different glue planes to control for potential fabrication asymmetry. None was found. **(C)** Fabricated dice shown left to right: classes 1-23, centered anisotropic dice, and high-contrast die for high-speed filming. ArUco codes reprinted from [12] under a CC BY license, with permission from Rafael Salinas, original copyright 2014.

A 24th die class was introduced to test the effects of a face-centered inertia tensor (Fig. 2D). In particular, the axis associated with the minimum principal moment was normal to die faces 3 and 4. For a given torque, rotation about this minimum-moment axis has the greatest angular acceleration. We therefore hypothesized that the remaining four faces might have increased landing probabilities. The two halves of this die can only be fabricated in a single glue plane; nevertheless, two replicates were tested, and each produced a distribution close to uniform; when pooled. However, the rolls showed a small but statistically detectable departure from uniformity (Fig. 2D and Fig. S2). An STL file is included as supplementary material.

## Automatic roller operation

The rolling platform (previous section) is mounted to a servo motor at the pivot of the seesaw. An Arduino sketch (see supplementary material) was used to automate the motion of the servo. During one complete cycle of the seesaw, the dice tumbled through the randomizer and down the ramp before settling in the landing area. After a pause for face detection, the dice were rocked backward to the beginning of the randomizer. A full cycle was completed in approximately 10 s. Throughout data collection, six dice (three replicates from each of two classes) were rolled simultaneously, yielding an average of 30,000–40,000 rolls in a 24-hour period. Data collection continued nightly at this rate for three weeks. Key components of our design are listed below:

- **ANNIMOS 45KG High Torque Servo 8.4V**
- **Pan Tilt Servo Mount Bracket for MG996R S3003**
- **arduino uno r3 Board**
- **Arducam 30fps@1080P HDR USB Camera Module**
- **Arducam VGA Global Shutter USB Camera Module, 800 FPS (see below)**

We acknowledge several limitations of our roller design. Occasionally, dice become trapped between randomizer elements and fail to complete rolls. Because this limitation is temporary and the rolls were independent, its only effect was a reduced rate of data collection. Surprisingly, dice occasionally become trapped in the landing area despite tilting the seesaw towards the randomizer to start the next roll cycle. In these cases, dice remain in the landing area and could be recorded repeatedly without rerandomization. To mitigate the effect on results, rolls were not recorded for dice that failed to change position or orientation within the landing area.

Sticker durability became an issue after many collisions with the randomizer elements. Peeling sticker corners and face wear interfered with roll detection, which depended on the clarity of the ArUco codes. This issue was mitigated by applying clear matte Mod Podge to dice faces prior to rolling. Vinyl stickers are more durable but were omitted from the current study to avoid altering the die surfaces. However, we acknowledge that the felt rolling surface of the seesaw may have changed during data collection. Conceivably, wear of the dice or rolling surface could affect the tumbling dynamics of dice during the course of data collection. This could explain the monotonic decrease in the fitted energy parameter (β) of the Boltzmann distributions (Fig. 2A). We addressed this possibility by reproducing similar fitted values of β with a second set of rolls for each die class after the accumulation of potential felt surface wear.

## Automated detection and counting of die-face transitions

High-speed videos were recorded using a monochrome global-shutter camera at a nominal resolution of 320 × 240 pixels and 240 frames $s^{-1}$. Automated face recognition was challenging because the visible die face typically occupied only approximately 10–20 pixels, underwent rapid translation and rotation, and was frequently affected by motion blur, perspective distortion, and changes in illumination. In addition, the frame-rate metadata stored in some AVI files did not reflect the acquisition rate; therefore, the known acquisition rate of 240 frames $s^{-1}$ was supplied explicitly during analysis. Preliminary tests with ArUco markers were unsuccessful because the markers were too small and blurred for reliable corner detection. Direct template matching was also sensitive to changes in orientation and image quality. We therefore marked the six faces with visually distinct, high-contrast grayscale symbols: a solid circle, ring, plus sign, filled triangle, three parallel bars, and a checker pattern.

A custom OpenCV labeling program was used to construct the training dataset[13]. For each selected video frame, a bounding box was placed around the visible die and the corresponding face was assigned a label from 1 to 6. The program saved both the complete 320 × 240 frame with a YOLO-format bounding-box annotation and a padded square crop centered on the die. Frames were generally sampled at intervals rather than labeling every frame because adjacent frames in the 240-frames $s^{-1}$ recordings were highly redundant. Nine videos were labeled in this manner. To prevent information leakage from nearly identical neighboring frames, the data were divided by complete video rather than by individual image: eight videos were used for training, one labeled video was used for validation, and an additional video was withheld for final testing.

Localization and face identification were treated as separate learning problems. For detector training, all original face-specific bounding-box labels were collapsed into a single class, *die*. This simplified the localization problem and allowed empty annotation files to serve as negative background examples. A pretrained Ultralytics YOLO26n detector was fine-tuned for 150 epochs using an input size of 960 pixels, a batch size of 8, mosaic augmentation disabled, and early stopping with a patience of 30 epochs[14,15]. The relatively large detector input size was used to preserve spatial information for the small die in the original frames. Face identity was learned using a separate YOLO26n classification model. The labeled die crops were organized into six classes, resized to 128 × 128 pixels, and used to fine-tune the classifier for 150 epochs with a batch size of 64 and an early-stopping patience of 30 epochs. Both models were trained with Ultralytics 8.4.96 and PyTorch 2.8.0 using an NVIDIA RTX A5000 GPU[14,15].

Recorded videos were subsequently analyzed offline with a custom two-stage processing program (supplementary video S1). For each frame, the detector identified candidate die locations, and only the highest-confidence detection was retained because each video contained a single die. The detected box was expanded by 25% on each side, converted to a square crop, and resized to the classifier input dimensions. The classifier returned probabilities for all six faces rather than only the most likely class. Optional test-time augmentation classified the crop at rotations of 0°, 90°, 180°, and 270° and averaged the resulting probability vectors, improving robustness to arbitrary die orientation at the cost of additional processing time.

Frame-by-frame classifications were not counted directly as transitions because occasional detection errors could otherwise produce rapid false changes between faces. Instead, the six face probabilities were smoothed using a centered, detector-confidence-weighted moving window. A face was treated as unknown when its probability was below the confidence threshold or when the difference between the highest and second-highest probabilities was insufficient. A new face state was confirmed only after it received support in at least three observations, while short gaps of up to two unknown frames were permitted between supporting observations. Because processing was performed after data collection, a confirmed state could be assigned retrospectively to the first supporting frame. A face transition was counted only when the confirmed stable state changed from one recognized face to another; transitions into or out of the unknown state were not counted.

For each recording, the analysis produced an annotated video showing the detected bounding box, predicted face, confidence, and cumulative number of transitions. It also produced a frame-level table containing the detector confidence, bounding-box coordinates, probabilities for all six faces, raw classification, smoothed classification, and confirmed stable state. Separate output files recorded the timing and identity of each confirmed transition and summarized the results as a 6 × 6 face-transition count matrix. This two-stage approach allowed localization, face classification, and temporal decoding to be evaluated independently and reduced false transition counts caused by brief motion blur, missed detections, or isolated classification errors.

## MuJoCO simulations

Rigid-body simulations were performed in MuJoCo using a custom Python program that reproduced the 23 rotation-distinct corner-loading classes[9]. Each die was represented as a 20-mm cube with a 4.0-g uniform base body and a 1.0-g point mass at each occupied corner, positioned 10 mm from the geometric center along each Cartesian axis. For each class, the total mass, center-of-mass, and full inertia tensor about the center-of-mass were calculated from the corresponding corner pattern and written directly to the MuJoCo inertia specification. This procedure preserved the same external geometry and contact parameters across classes while allowing their mass properties to vary. A total of 6,000 accepted rolls was simulated for each class, yielding 138,000 simulated outcomes.

Each roll began from rest with zero linear and angular velocity and a uniformly randomized three-dimensional orientation (supplementary video S2). The die position was adjusted for its class-specific center-of-mass offset so that the center-of-mass began at the same world height, z = 0.700 m, for every class and orientation; the release coordinates were x = −0.23 m and y = 0 m. Initial-condition seeds were shared by roll index across die classes, allowing classes to be compared under matched randomized release orientations. The MuJoCo integration timestep was 0.0005 s.

The simulated rolling environment contained a fixed ramp inclined at 45°, with a length of 1.7 m, width of 0.30 m, and sliding-friction coefficient of 1.2. A staggered pegboard containing six rows and six columns of cylindrical obstacles occupied the upper portion of the ramp and introduced repeated collisions and changes in orientation. The landing region extended from −0.05 to 0.20 m along the x-axis and from −0.15 to 0.15 m along the y-axis. Physical tray walls 0.080 m high and 0.006 m thick surrounded the landing region, with the negative-x side left open to receive dice from the ramp. The ramp, pegboard, landing boundaries, and class-specific inertia properties were generated programmatically from a common MuJoCo XML template.

A die was classified as settled when its linear speed remained below 0.02 m $s^{-1}$ and its angular speed remained below 0.20 rad $s^{-1}$ for at least 0.25 s. Simulations were terminated after settling or after a maximum simulated duration of 2.0 s. The upward face was identified from the body-fixed face normal having the greatest alignment with the world vertical axis. For transition analysis, a face state was recorded only when its vertical alignment was at least 0.75 and remained stable for at least 0.025 s. Translational and rotational kinetic energies were sampled every 0.01 s. A roll was retained only if the die settled within the defined landing region; unsuccessful attempts were repeated with a new deterministic random seed, up to 20 attempts per requested roll. In the production run of 6,000 rolls per class, the simulation results were stored directly for subsequent analysis rather than rendered as individual videos.

## Six-state dissipation model

The six-state model was implemented in Python using NumPy for Monte Carlo simulation, pandas for tabulation of results, and Matplotlib for visualization[16–18]. Each state represented one possible upper face of a three-corner-loaded die, with state values assigned from the corresponding corner advantages. Cube-face adjacency was encoded explicitly so that transitions could occur only between neighboring faces. For each dissipation rate, many independent trajectories were simulated in parallel using vectorized NumPy operations. Initial states were randomized. At each time step, a neighboring state was selected randomly, the proposed transition was accepted according to the difference in corner advantage and the remaining time-dependent mechanical energy, and the energy scale was reduced exponentially to represent dissipation. The final occupied state was recorded for each trajectory, and the resulting frequencies were used to estimate final-state probabilities as a function of dissipation rate. The model was intended as a qualitative demonstration of dissipative trapping rather than a calibrated simulation of the physical rolling apparatus.

## Statistical analysis

Statistical tests were selected to match the structure of each analysis. For Figure 2A, the monotonic relationship between the number of filled corners and the fitted sensitivity parameter, $\beta^n$ , was evaluated using a Spearman rank correlation. Because the hypothesized direction was a decrease in $\beta_n$ with increasing corner filling, significance was assessed using a one-sided test. For Figure 2C, the association between normalized mean tipping resistance and the mean landing probability of zero-advantage faces was evaluated using a Spearman rank correlation, with each four-corner-loaded die class treated as one independent observation. For Figure 2D, the pooled face counts from the balanced anisotropic die were compared with the equal-frequency expectation of 1/6 per face using a chi-square goodness-of-fit test. For Figure 4, the distributions of transition count, transition rate, settling time, and centroid path length were compared between unfilled and fully corner-loaded dice using two-sample Kolmogorov–Smirnov tests. The four Figure 4 comparisons were treated as a single family, and their p-values were adjusted using the Holm procedure. Unless otherwise stated, tests were two-sided, and $p<0.05$ was considered statistically significant.

## Data Availability

We used freely available python packages to perform our analysis. Data and code used for manuscript preparation are freely available online: nkinney06/corner_loaded_dice.

## Acknowledgements

We thank several students who participated in dice construction and labeling of training data: Emma Evans, Emma Leahy, Emily Kempin, Caitlyn Uimari, and Maggie Booth. We thank Mona Browning for helpful suggestions about the design of randomizer elements used for the automatic roller apparatus. We thank Kirk Kinney for assistance with design and construction of the roller apparatus.

## Citations

# Supporting information captions

**Figure S1. Observed (top) and Simulated (bottom) landing probabilities for the 23 classes of corner-loaded dice.** The simulated landing probabilities are consistent with experiment.

**Figure S2 – Observed (left) and Simulated (right) landing probabilities for the face-centered anisotropic die.** The discrepancy from experiment is beyond the scope of this study but may reflect idealized contact and friction parameters, the point-mass approximation used to represent the internal weights, or differences between the simulated and physical release and landing conditions.